\documentclass[letter]{aa}

\usepackage{graphicx}
\usepackage{caption}
\usepackage{subcaption}
\usepackage{epstopdf}
\usepackage{amsmath}
\usepackage[english]{babel}
\usepackage{amssymb}
\usepackage{wrapfig}
\usepackage{pdflscape}
\usepackage{lscape}
\usepackage{longtable}
\usepackage{appendix}
\usepackage{textcomp}
\usepackage{multirow}
\bibpunct{(}{)}{;}{a}{}{,}
\usepackage{array}
\usepackage{ragged2e}
\usepackage{txfonts}
\begin{document} 

   \title{Radio jets and gamma-ray emission in radio-silent narrow-line Seyfert 1 galaxies \thanks{37~GHz data are only available in electronic form at the CDS via anonymous ftp to cdsarc.u-strasbg.fr (130.79.128.5) or via http://cdsweb.u-strasbg.fr/cgi-bin/qcat?J/A+A/
}}

 %  \subtitle{}

   \author{A. L\"{a}hteenm\"{a}ki\inst{1}\fnmsep\inst{2}\thanks{\email{anne.lahteenmaki@aalto.fi}}
          \and
          E. J\"{a}rvel\"{a}\inst{1}\fnmsep\inst{2}
          \and
          V. Ramakrishnan\inst{1}\fnmsep\inst{3}
          \and
          M. Tornikoski\inst{1}
          \and
          J. Tammi\inst{1}
          \and
          R. J. C. Vera\inst{1}\fnmsep\inst{2}
          \and
          W. Chamani\inst{1}\fnmsep\inst{2}
          }

   \institute{Aalto University Mets\"{a}hovi Radio Observatory, Mets\"{a}hovintie 114, FI-02540 Kylm\"{a}l\"{a}, Finland
         \and
             Aalto University Department of Electronics and Nanoengineering, P.O. Box 15500, FI-00076 AALTO, Finland
        \and
             Astronomy Department, Universidad de Concepci\'on, Casilla 160-C, Concepci\'on, Chile
             }

   \date{Received ; accepted }

% \abstract{}
% 5 {} token are mandatory

  \abstract
   { %context
   We have detected six narrow-line Seyfert 1 (NLS1) galaxies at 37~GHz that were previously classified as radio silent and two that were classified as radio quiet. These detections reveal the presumption that NLS1 galaxies labelled radio quiet or radio silent and hosted by spiral galaxies are unable to launch jets to be incorrect. The detections are a plausible indicator of the presence of a powerful, most likely relativistic jet because this intensity of emission at 37~GHz cannot be explained by, for example, radiation from supernova remnants. Additionally, one of the detected NLS1 galaxies is a newly discovered source of gamma rays and three others are candidates for future detections.
   }
   

   \maketitle

%________________________________________________________________

\section{Introduction}
\label{sec:intro}

It has been conventionally assumed that relativistic jets can be exclusively launched by highly evolved elliptical galaxies harbouring supermassive central black holes of masses larger than $10^{8} M_{\odot}$. Bright radio emission, and in some cases gamma-rays as well, are generated in these powerful jets propagating at almost the speed of light. In contrast, narrow-line Seyfert 1 (NLS1) galaxies are presumably young active galactic nuclei (AGN) \citep{2001mathur1} with supermassive black holes of intermediate mass ($M_{\textrm{BH}} < 10^{8} M_{\odot}$)\citep{2000peterson1}, high accretion rates \citep{1992boroson1}, and preferably spiral-type host galaxies, even though a small percentage of elliptical and interacting or disturbed systems have also been identified \citep{2007ohta1,2007zhou1,Anton2008,2014leontavares1,Olguin-Iglesias2017,2017dammando1,2018dammando1}. NLS1 galaxies are usually classified as radio-quiet or completely radio-silent sources, and in most cases the faint radio emission is assumed to originate in star formation processes. Furthermore, late-type galaxies are typically not supposed to be able to launch powerful relativistic jets. However, compact jetted radio morphologies typical of young AGN have occasionally been seen \citep{2010gliozzi1,2015doi1,2015richards1,2015gu1,Lister2016,Congiu2017} in a minority of NLS1 galaxies that, in contrast to Seyfert galaxies in general, resemble blazar-type AGN in their properties. Less than 20 gamma-ray emitting NLS1 galaxies have been detected by the \emph{Fermi} satellite \citep{2009abdo1,2009abdo2,2012dammando1,2018paliya1}: these are all radio loud as expected of powerful jetted sources, which, however, is uncharacteristic of NLS1 galaxies.

Regardless of how different in their initial properties NLS1 galaxies may be from the more energetic AGN, the radio behaviour of the radio-loud NLS1 galaxies is similar to the blazar-type AGN showing considerable variability over various timescales \citep{2017lahteenmaki1}. 
However, radio-loud NLS1 galaxies, consisting of only around 10\% of all NLS1 sources, are not representative of the whole source class but are rather a distinct minority. An even smaller fraction of these galaxies appear to possess jets. Moreover, the radio loudness parameter is defined using one-epoch, low-frequency (usually 1.4 or 5~GHz) observations that give no insight into, for example, variability over time. 

Studying an unprejudiced sample of sources that is not restricted by, for example, the precarious concept of radio loudness \citep{Padovani2016,2017jarvela1, 2017padovani1}, which is dependent on observing frequency and epoch, is key in examining the genuine diversity ---reflected in their intrinsic and environmental properties--- of the NLS1 population. For example, searches for gamma-ray emission in NLS1 galaxies often concentrate only on sources that are classified as radio loud and therefore exhibit blazar-type behaviour. As successful as these searches seem at first glance, this approach also omits many sources. To conquer these issues we selected samples of NLS1 galaxies classified as radio quiet and radio silent and observed them at 37~GHz to look for detections and variability that would indicate the presence of radio jets.

\section{Sample and observations}

In our previous study of 78 of the most radio-loud NLS1 galaxies at 37~GHz \citep{2017lahteenmaki1}, we found the detection rate to be 19\%. In this study we included 66 NLS1 galaxies of lower or absent radio loudness parameters from radio-silent ---without counterparts in the Very Large Array Faint Images of the Radio Sky at Twenty-Centimeters (VLA FIRST) survey--- to radio-quiet sources. We based the selection on features such as extraordinarily dense large-scale environment \citep{2017jarvela1} or their multiwavelength properties, making these galaxies promising candidates for high-frequency radio observations.

In the first case we selected 25 sources with the highest densities as it is assumed that the large-scale environment affects the evolution of the galaxy and therefore possibly also jet launching \citep{2017jarvela1}. The four sources in this study that reside in superclusters belong to this sample. In the latter case we picked 41 sources based on two criteria. First, we evaluated whether the 37~GHz flux, which were extrapolated by eye from their spectral energy distributions (SEDs) in the Space Science Data Center (SSDC), has the potential of exceeding our detection limit; and alternatively, we required that their optical or X-ray emission levels in the NASA/IPAC Extragalactic Database (NED) are unusually high compared to their archival radio emission level, implicating that they could occasionally be detected at high radio frequencies. The four remaining sources are included in this sample. The optical and X-ray emission levels of the sources chosen based on their large-scale environment are on average lower than those of the sources selected for their SEDs, indicating that a higher optical and X-ray emission level is not a determining factor when predicting the detectability of NLS1 galaxies at high radio frequencies. The average number of observations presented here is 40.8 for each source, varying between 21 and 71, with a median value of 34. The average number of detections per source is 5.9, with minimum, maximum, and median values of 1, 15, and 3.5, respectively. The 37 GHz observations are described in appendix~\ref{sec:37ghzdata}.

\section{Results}
 
We have detected eight of these unconventionally selected sources at 37~GHz, the detection limit being of the order of 0.2~Jy under optimal conditions. The detection rate is 12\%, which is unexpectedly close to the detection rate of the earlier sample, originally defined as radio loud. This shows that powerful, most likely relativistic jets are not exclusively launched by blazar-type NLS1 galaxies but seem to be the property of the classic Seyfert-type as well ---albeit in a more modest scale. 

We also report the discovery of gamma-ray emission from one of our sources (maximum-likelihood test statistic, TS $>$ 25): J164100.10+345452.7. This source is located 0.10{\textdegree} from the radio position. The TS value of 39 confirms this source as a new gamma-ray emitting NLS1 galaxy. Its radio loudness of 13 makes it a borderline source between radio loud and radio quiet, however, it has been detected at 37~GHz twice. The 37~GHz radio flux is fairly modest with peak flux density of 0.46~Jy, positioning it at the low end of the radio versus gamma-ray emission continuum. The source resides in a large-scale environment that is on a denser side of intermediate \citep{2017jarvela1}. 
Another three sources are prospective gamma-ray emitting candidates (9 $<$ TS $<$ 25): J090113.23+465734.7, J122844.81+501751.2, and J123220.11+495721.8. The \emph{Fermi}--LAT data analysis is described in appendix~\ref{sec:fermidata}.
NLS1 galaxies exhibiting gamma-ray emission typically show similar radio behaviour as gamma-ray-bright AGN. Some of these galaxies seem to imitate BL Lac objects (BLOs); most of the time they are barely detected at 37~GHz, yet they can be bright at gamma rays \citep{2003lahteenmaki1,2015foschini1}. Examples of such BLOs are Mark~421 and ON~231. At the other end there are the flat-spectrum radio-loud quasars (FSRQs) with high radio and gamma-ray emission levels. While the jet powers of NLS1 sources are comparable to FSRQs, they have even lower black hole masses than BLOs \citep{2015foschini1}, setting them apart from the blazar-type AGN. 

\begin{figure*}[ht!]
    \centering
        \begin{subfigure}[t]{0.49\textwidth}
        \centering
        \includegraphics[height=5.4cm]{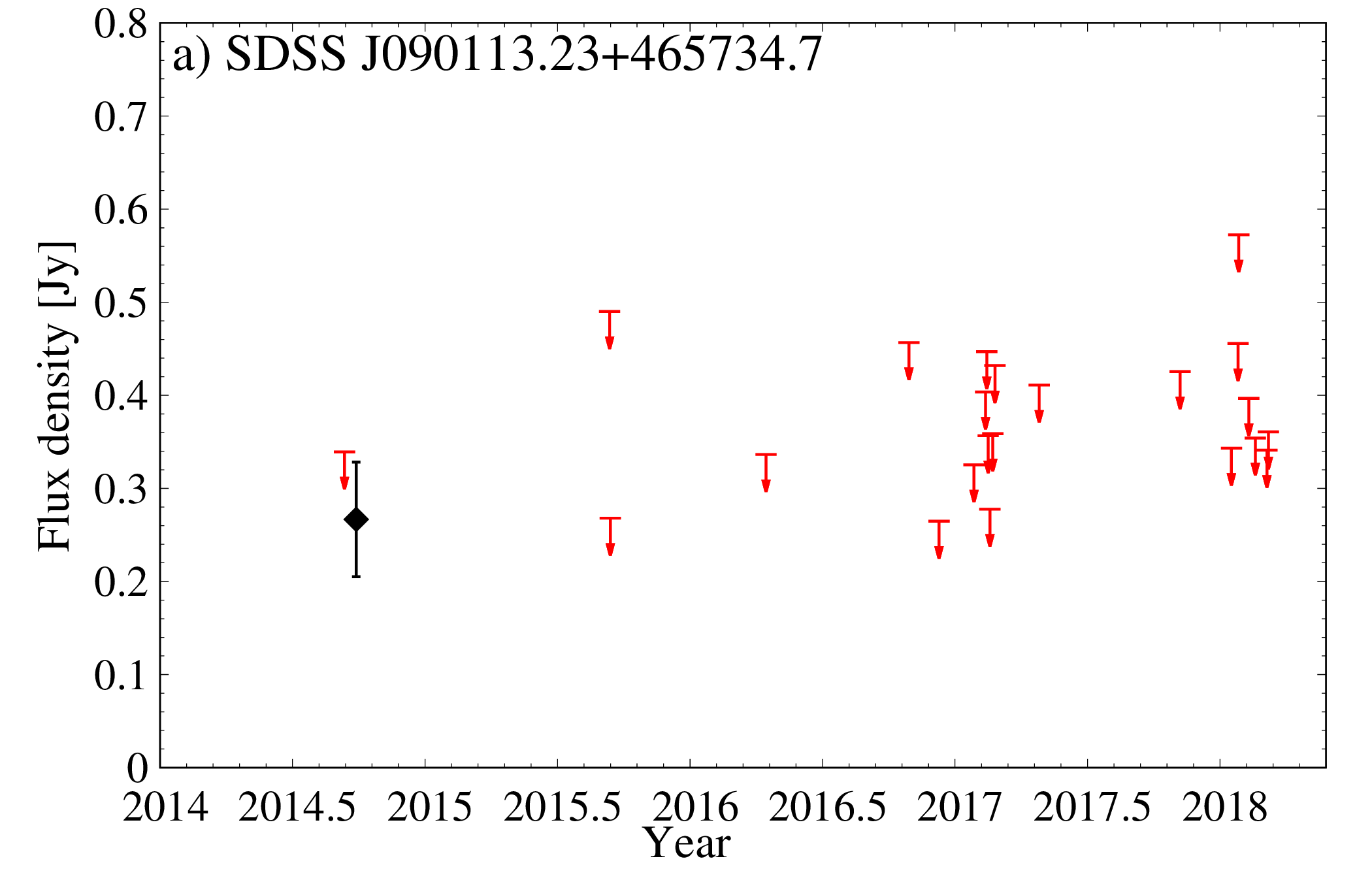}
     \label{fig:lcJ0901}
    \end{subfigure}%
    ~ 
    \begin{subfigure}[t]{0.49\textwidth}
        \centering
        \includegraphics[height=5.4cm]{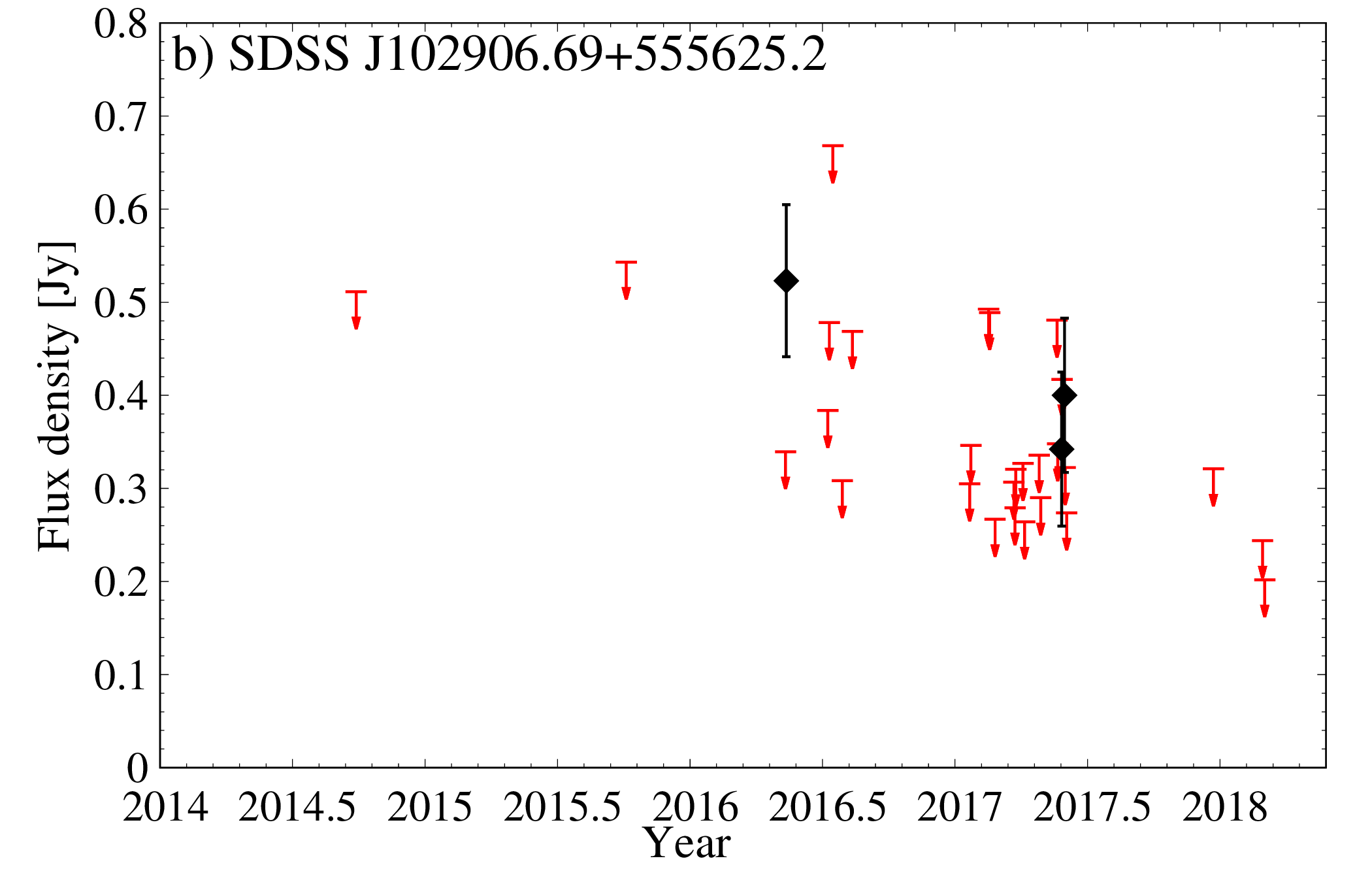}
         \label{fig:lcJ1029}
    \end{subfigure}
    ~\\
     \begin{subfigure}[t]{0.49\textwidth}
        \centering
        \includegraphics[height=5.4cm]{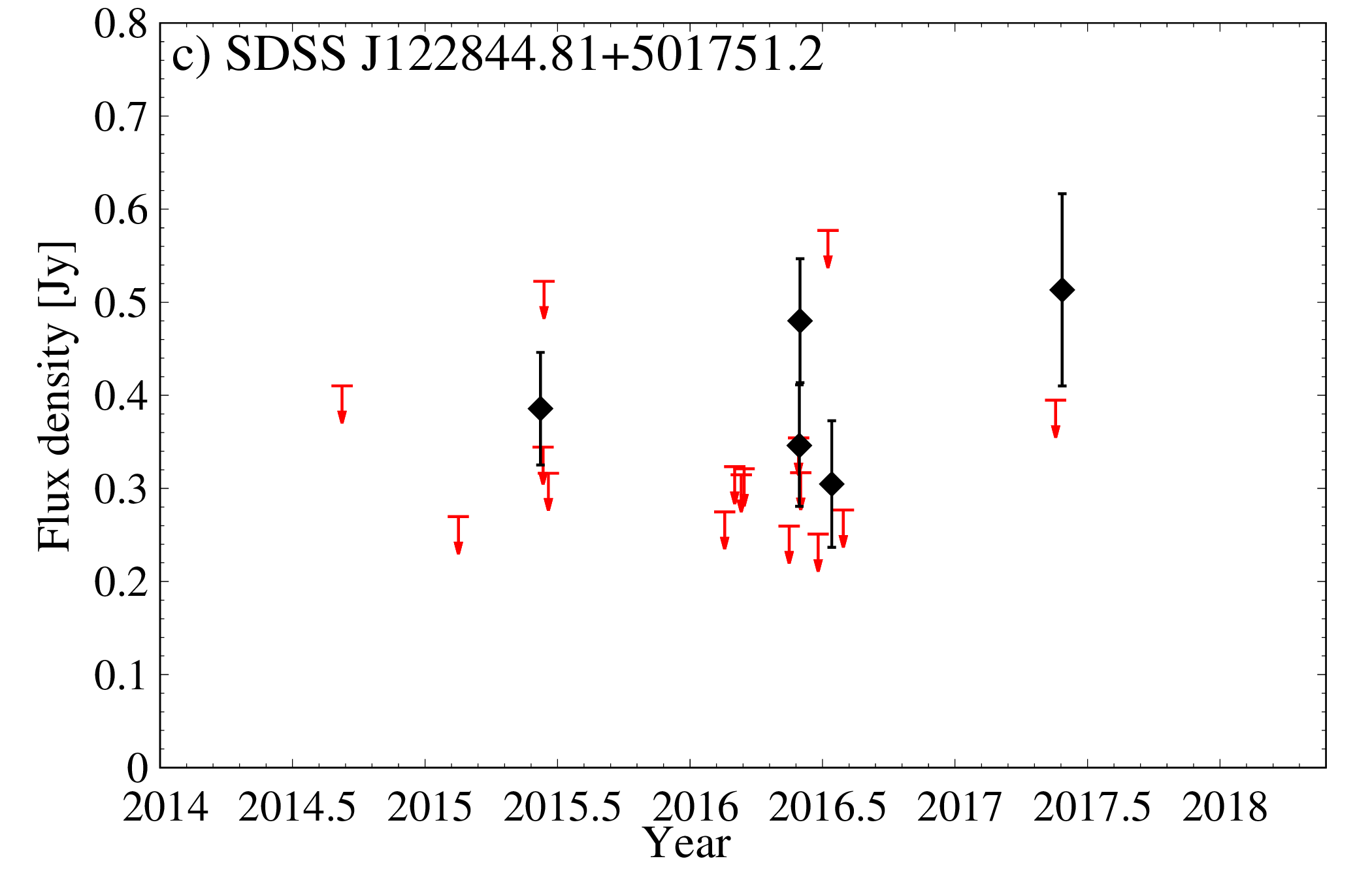}
         \label{fig:lcJ1228}
    \end{subfigure}
    ~
    \begin{subfigure}[t]{0.49\textwidth}
        \centering
        \includegraphics[height=5.4cm]{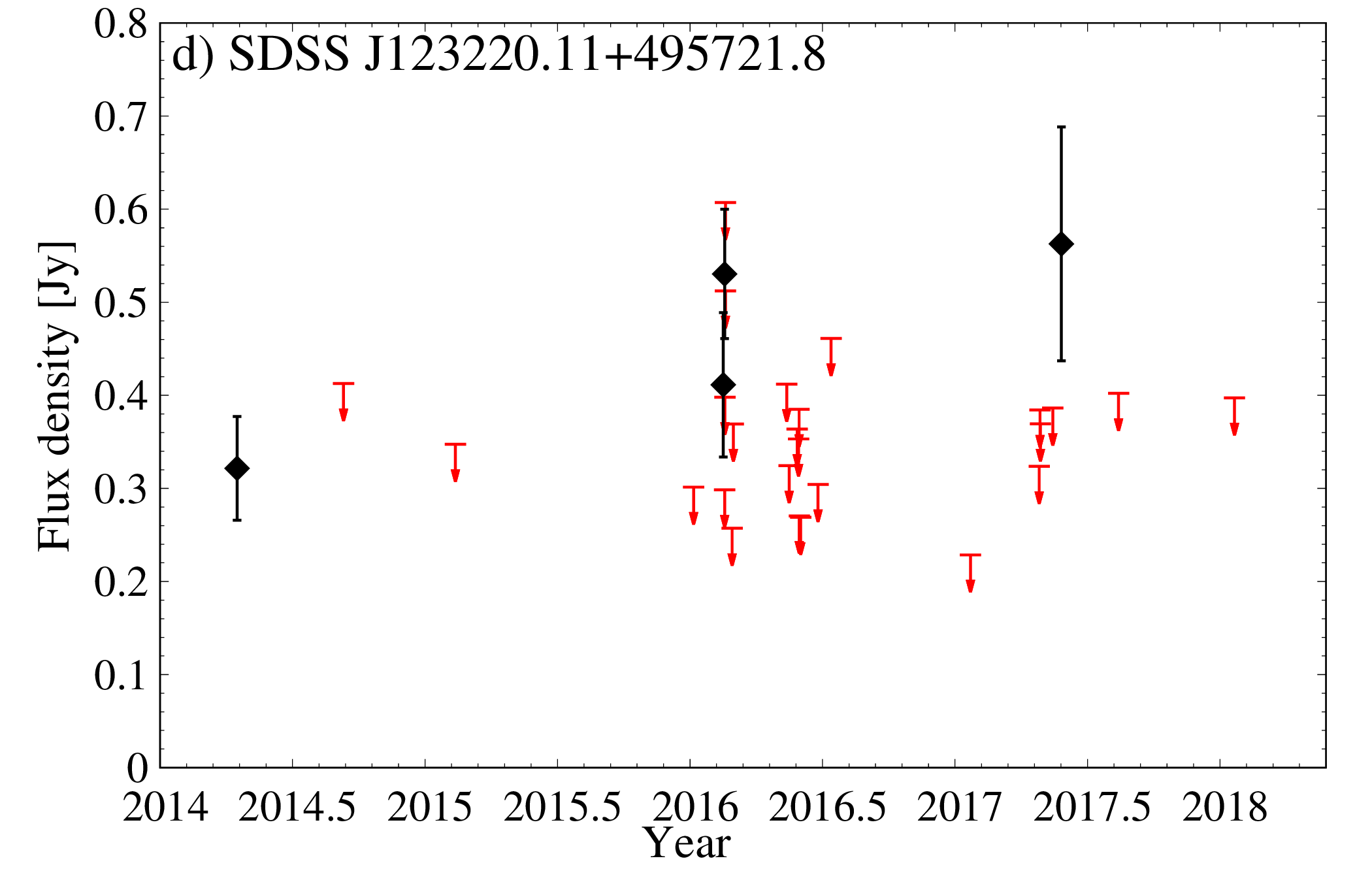}
         \label{fig:lcJ1232}
    \end{subfigure}
    ~\\
    \begin{subfigure}[t]{0.49\textwidth}
        \centering
        \includegraphics[height=5.4cm]{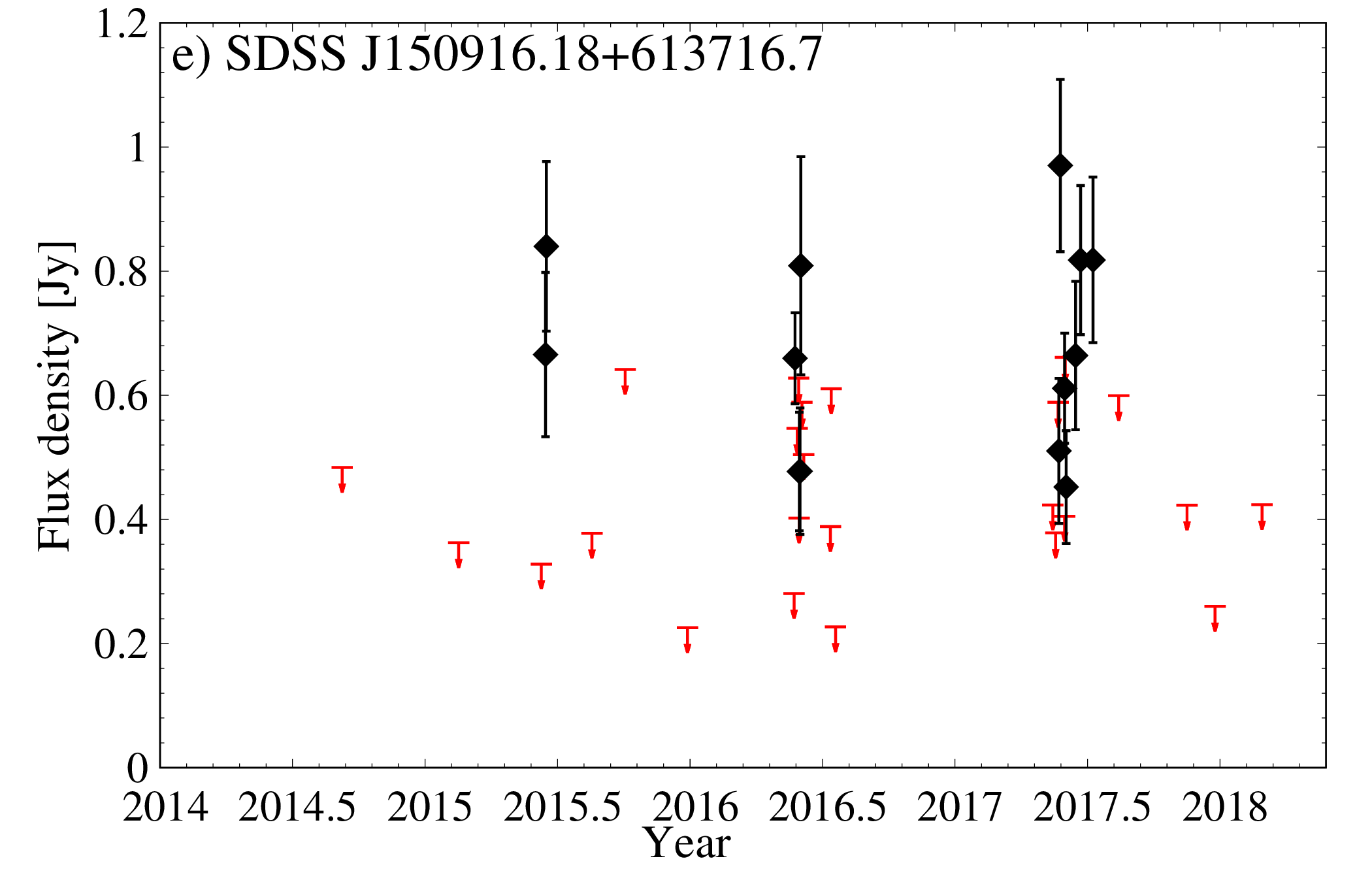}
        \label{fig:lc1509}
    \end{subfigure}%
    ~ 
    \begin{subfigure}[t]{0.49\textwidth}
        \centering
        \includegraphics[height=5.4cm]{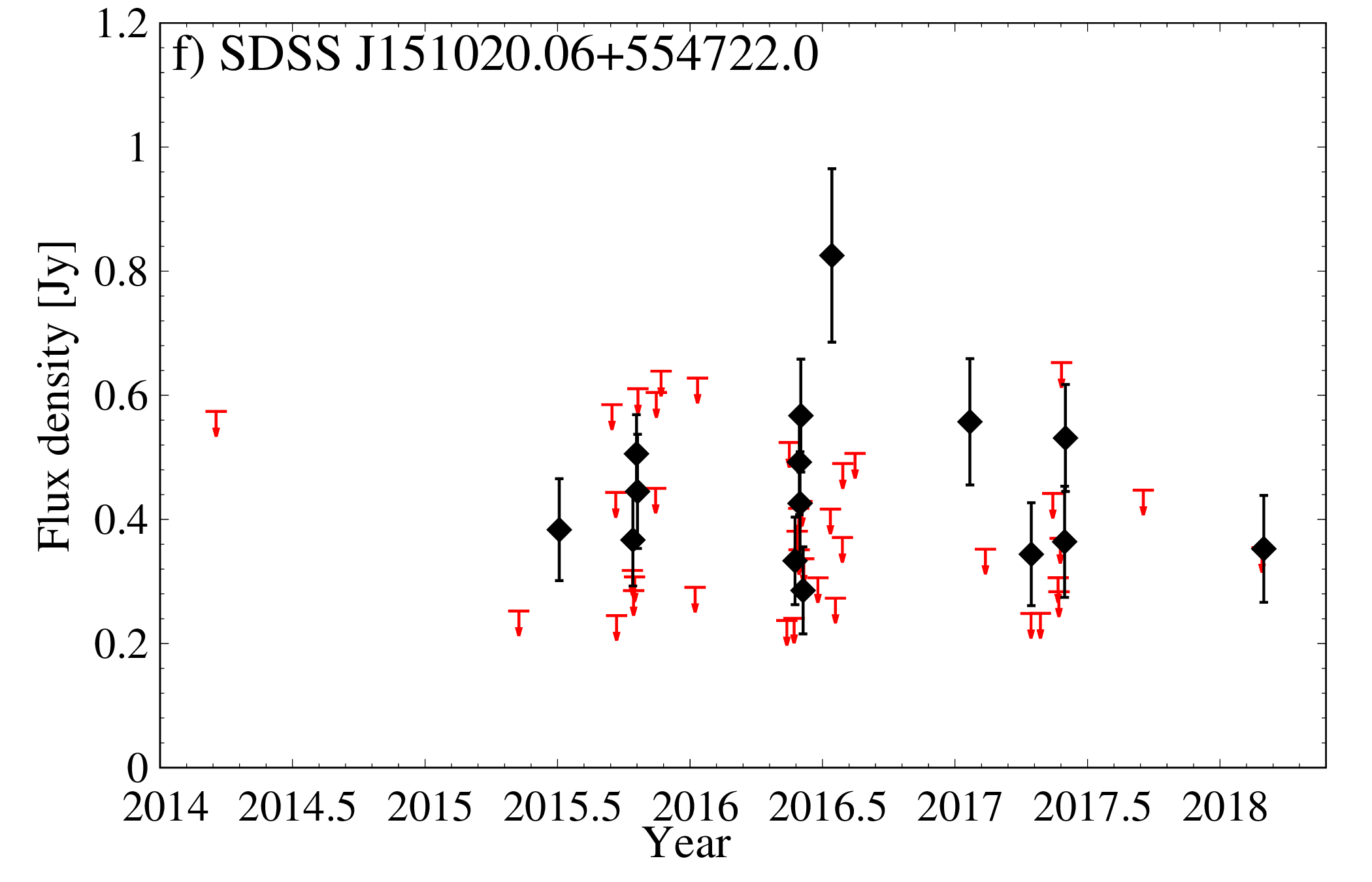}
        \label{fig:lcJ1510}
    \end{subfigure}
    ~\\
     \begin{subfigure}[t]{0.49\textwidth}
        \centering
        \includegraphics[height=5.4cm]{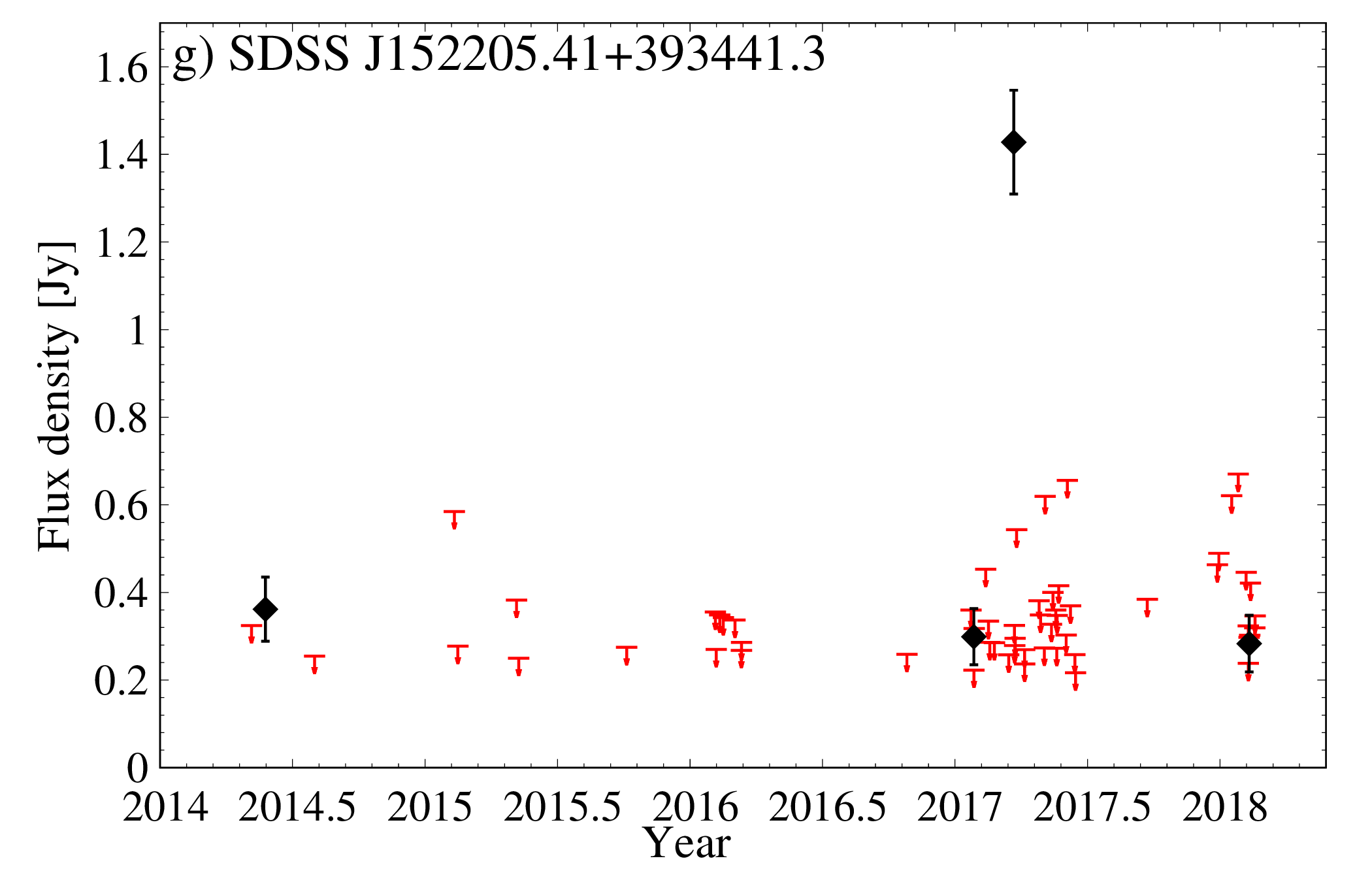}
         \label{fig:lcJ522}
    \end{subfigure}
    ~
    \begin{subfigure}[t]{0.49\textwidth}
        \centering
        \includegraphics[height=5.4cm]{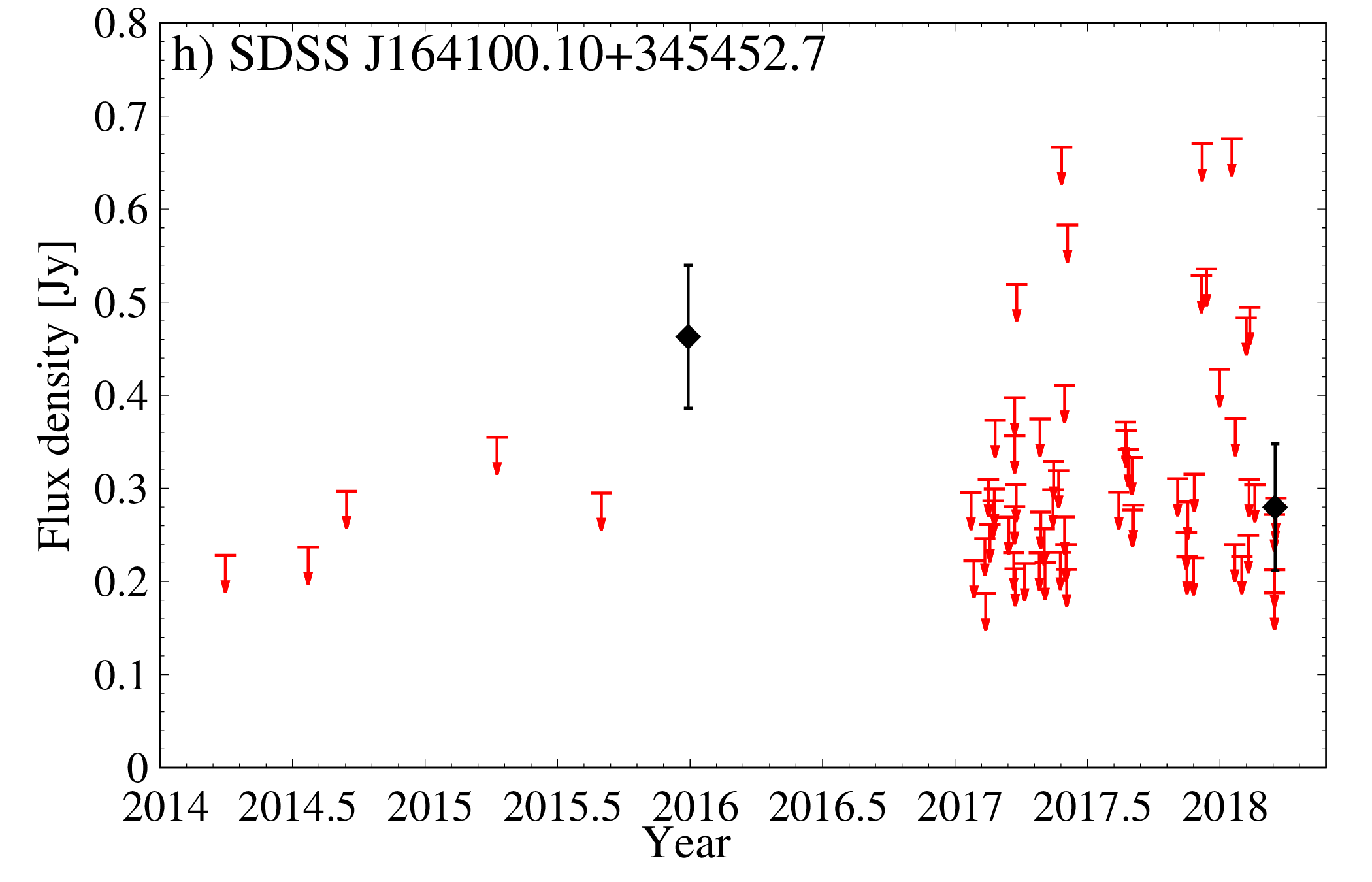}
     \label{fig:lcJ641}
    \end{subfigure}
    \caption{Flux density curves at 37~GHz. Detections and their errors are denoted by black diamonds, and four-sigma upper limits as red arrows.} \label{fig:firstfour}
\end{figure*}

A summary of 37~GHz radio and gamma-ray data is shown in Table~\ref{tab:radiogamma}. Column 1 gives the name of the source and Col. 2 its redshift.  The average flux density at 37~GHz is shown in Col. 3, and the gamma-ray flux and its error plus upper limits in Col 4. The TS value of the gamma-ray emission is given in Col. 5 (detection and candidates in bold). The large-scale environment of the source \citep{2017jarvela1} is defined in Col. 6 and Col. 7 gives the black hole mass either taken from \citet{2015jarvela1} for the radio-quiet sources or estimated as described in appendix \ref{sec:mbh} for the radio-silent sources. Parameter $R$ ($S_{\textrm{1.4~GHz}} / S_{\textrm{440~nm}}$), defining the radio loudness, is listed in Col. 8 \citep{2015jarvela1}. Radio flux density curves of the sources are shown in Figure~\ref{fig:firstfour}. Detailed 37~GHz statistics are available in Table~\ref{tab:mh}.

\begin{table*}[ht]
\caption[]{{Basic data and statistics of 37~GHz and gamma-ray observations.}} 
\centering
\begin{tabular}{l l l l l l l l}
\hline\hline
Source                   &  $z$  & $\mathrm{S}_{\text{37GHz,ave}}$ & $\mathrm{S}_{\gamma}$ & TS         & large-scale  & log $M_{\text{BH}}$ & $RL$ \\
                         &       & (Jy)                            & (10$^{-9}$ ph cm$^{-2}$ s$^{-1}$) &     & environment  &  ($M_{\odot}$)  & \\ \hline
SDSS J090113.23+465734.7 & 0.430 & 0.27                            & < 1.80  & {\bf 10}      & supercluster &  7.15               & -- \\
SDSS J102906.69+555625.2 & 0.451 & 0.42                            & < 0.97  & 1         & supercluster &  7.33               & -- \\
SDSS J122844.81+501751.2 & 0.262 & 0.41                            & < 6.59  & {\bf 20} & supercluster &  6.84               & -- \\
SDSS J123220.11+495721.8 & 0.262 & 0.46                            & < 1.39  & {\bf 16} & supercluster &  7.30               & -- \\
SDSS J150916.18+613716.7 & 0.201 & 0.67                            & < 2.24  & 1        &    void      &  6.66               & -- \\
SDSS J151020.06+554722.0 & 0.150 & 0.45                            & < 1.47  & --       & intermediate &  6.67               & -- \\
SDSS J152205.41+393441.3 & 0.077 & 0.59                            & < 1.44  & --       & void         &  5.97               & 2 \\
SDSS J164100.10+345452.7 & 0.164 & 0.37                            & 12.5 $\pm$ 2.18 & {\bf 39} & intermediate &  7.15               & 13 \\ \hline

\end{tabular}
\label{tab:radiogamma}
\end{table*}

A previous study by \citet{2017jarvela1} found a tendency for radio-silent NLS1 galaxies to mainly lie in voids and intermediate-density regions of the large-scale environment, in contrast to radio-loud sources preferably located in superclusters. However, a significant fraction of radio-silent NLS1 sources are located in superclusters as well \citep{2017jarvela1}. Of the eight sources presented here, four are found in voids and intermediate-density regions and four in superclusters. The two most radio-bright sources (average 37~GHz flux density $\sim$0.7~Jy) and the new gamma-ray source are, perhaps against expectations, all located in the most diffuse regions. Three of the sources in the most diffuse areas also possess the lowest black hole masses. These are all close by, matching the scenario in which older, more evolved galaxies (such as, for example, radio galaxies) are positioned at higher redshifts, and younger, more active galaxies (for example, quasars) at lower redshifts. Even though the differences between the redshifts of our eight sources are not that large and the sample is small, the tendency is clear. This highlights the problem in defining NLS1 samples correctly: radio emission is detected from sources defined as radio silent, and they are located in vastly disparate large-scale environments, indicating profoundly different stages of evolution.

Host galaxy information of NLS1 sources is rare and available only for three of our sources. Of these, J151020.06+554722.0 is a spiral galaxy and J123220.11+495721.8 possibly a spiral. J152205.41+393441.3 is an interacting system of two galaxies in which the NLS1 source is a spiral galaxy and the companion an unidentified, non-active galaxy; the two galaxies are located at identical redshifts \citep{2018jarvela1-host}.

\section{Discussion}

All of our sources reside in an area of the sky that is covered by the FIRST survey at 1.4~GHz, complete down to $\sim$1~mJy. Two of the sources were detected in the survey at flux densities below 5~mJy, but the remaining six sources were not, indicating 1.4~GHz flux densities at sub-mJy levels. However, detections at 37~GHz reveal that these sources exhibit jets and substantial activity in the nucleus. The FIRST observations were performed approximately 20 years ago. It is possible that these sources have been in an inactive radio state when observed -- although a non-relativistic jet might still have been present -- and thus classified as radio silent or radio quiet \citep{2009mundell1}. The activity may have since increased to a point where we can now detect the jet at 37~GHz. Another possibility is that they did not possess jets and were radio silent two decades ago, and we are now witnessing the triggering of a jet powerful enough to be detected at 37~GHz. Since no detectable relic radio emission exists in these sources, this might be an indication of short-timescale intermittent activity at radio frequencies \citep{2009czerny1}. The third option is that their radio spectra are inverted. In this case the non-simultaneous spectral indices (defined as $S_\nu \propto \nu^{\alpha}$) between 1.4 and 37~GHz for the radio-silent sources are between 1.7 and 2.1 for the lowest (0.27~Jy) and highest (0.97~Jy) 37~GHz flux densities, respectively, assuming a flux density of 1~mJy at 1.4~GHz. For J164100.10+345452.7 the indices are $\alpha_{\textrm{min}}$=1.4 and $\alpha_{\textrm{max}}$=1.6, and for J152205.41+393441.3 $\alpha_{\textrm{min}}$=1.4 and $\alpha_{\textrm{max}}$=1.9. These are crude estimates due to the temporal difference of the observations and possible variability of the 1.4~GHz emission. However, assuming that this scenario is true, the indices highlight the extreme nature of these sources.

Observations at lower frequencies, especially very long baseline interferometry observations necessary for examining the jet morphologies, are needed to discriminate between these scenarios and to confirm the presence of the relativistic jets. Such observations are forthcoming. The AGN phenomenon is clearly more complicated than previously assumed; sources able to develop powerful and most likely relativistic jets are found in diverse environments from voids to superclusters, have versatile selection of intrinsic properties, and most importantly, are not exclusively classified as radio loud and can be hosted by spiral galaxies. Using conventional radio loudness, determined with single-epoch low-frequency observations, as an indicator of jet activity or potential gamma-ray emission is not only useless but detrimental to unprejudiced studies of NLS1 galaxies, and therefore possibly also of other, particularly fainter AGN types.\\

\begin{acknowledgements}
This publication makes use of data obtained at Mets\"{a}hovi Radio Observatory, operated by Aalto University, Finland. Part of this work is based on archival data provided by the Space Science Data Center - ASI. This research has made use of the NASA/IPAC Extragalactic Database (NED) which is operated by the Jet Propulsion Laboratory, California Institute of Technology, under contract with the National Aeronautics and Space Administration.
\end{acknowledgements}

\bibliographystyle{aa}
\bibliography{artikkeli.bib}

\begin{appendix}

\section{37 GHz observations}
\label{sec:37ghzdata}

The 13.7 m radio telescope at Aalto University Mets\"{a}hovi Radio Observatory in Finland is used for monitoring large samples of AGN at 22 and 37~GHz. Our NLS1 observing programme, launched in 2012, currently consists of 185 NLS1 galaxies frequently monitored at 37~GHz. A detailed description of the programme, source selection, and the first results and data can be found in \citet{2017lahteenmaki1}. The measurements were made with a 1~GHz band dual beam receiver centred at 36.8~GHz. The observations are on--on observations, alternating the source and the sky in each feed horn. A typical integration time to obtain one flux density data point of a faint source is between 1600 and 1800~s. The sensitivity is limited by skynoise, and the results do not significantly improve after the used maximum integration time of 1800~s. The detection limit of our telescope at 37~GHz is of the order of 0.2~Jy under optimal conditions. Data points with a signal-to-noise ratio below four are handled as non-detections. Non-detections may occur either because the source is too faint or, for example, because of non-ideal weather. The flux density scale is set by observations of DR~21. Sources NGC~7027, 3C~274, and 3C~84 are used as secondary calibrators. A detailed description of the data reduction and analysis is given in \citet{terasranta98}. The error estimate in the flux density includes the contribution from the measurement rms and the uncertainty of the absolute calibration. Observation statistics are available in Table~\ref{tab:mh}. Columns 1 -- 3 give the name and the coordinates of the sources (in J2000). The number of detections and observations at 37~GHz is shown in Col. 4. and the detection percentage in Col. 5. Cols. 6 and 7 give the maximum flux density at 37 GHz and the 1.4 GHz flux density from FIRST. A table containing 37~GHz flux densities and their errors is available at the CDS.

We also performed test observations of a source with a descrepancy in its classification. In \citet{2015jarvela1} J090113.23+465734.7 is listed as radio loud. A search radius of 1 arcmin was used to obtain the radio data, and while there is no detected radio source at the source coordinates (the mean epoch of the FIRST observations is 1997.2), there is a radio source in its close proximity at coordinates RA 09h01m09.225s, Dec +46d57m21.19s. This is MJV~09737, which has no listed redshift or observations at other wavelengths. Their separation is 0.72 arcmin and it is therefore probable that the radio data in \citet{2015jarvela1} originate in this source, and J090113.23+465734.7 is apparently radio silent. The source has been detected at 37~GHz only once, and the detection raised concern about the possible contamination by MJV~09737 since it also lies within our antenna beam (2.4 arcmin at 37~GHz). The flux density of MJV~09373 at 1.4~GHz is 1.55~mJy and it has not been detected at higher radio frequencies; thus it is likely to be faint at 37~GHz. To ensure this we ran a set of simple tests to check how another radio source lying in the antenna beam might affect our observations, by observing the sources in pairs several times. All observations of MJV~09737 were non-detections so its contribution to 37~GHz observations seems negligible, but this test is not precise enough to certainly rule out possible contamination.

\begin{table*}[ht]
\caption[]{{Basic data and statistics of the 37~GHz observations.}} 
\centering
\begin{tabular}{l l l l l l l l}
\hline\hline
Source                   & RA            & Dec           & $N_{\textrm{Det}}$/$N_{\textrm{Obs}}$ & Det  & $S_{\text{37GHz, max}}$ & $S_{\text{1.4GHz}}$ \\
                         & (hh mm ss.ss) & (dd mm ss.ss) &                                       & (\%) & (Jy)                    &  (mJy)              \\ \hline
SDSS J090113.23+465734.7 & 09 01 13.23   & +46 57 34.67  & 1/23                                  & 4.3  & 0.27                    & --                 \\
SDSS J102906.69+555625.2 & 10 29 06.69   & +55 56 25.29  & 3/31                                  & 9.7  & 0.52                    & --                  \\
SDSS J122844.81+501751.2 & 12 28 44.78   & +50 17 51.31  & 5/21                                  & 23.8 & 0.51                    & --\tablefootmark{a}                 \\
SDSS J123220.11+495721.8 & 12 32 20.11   & +49 57 21.75  & 4/29                                  & 13.8 & 0.56                    & --                 \\
SDSS J150916.18+613716.7 & 15 09 16.17   & +61 37 16.80  & 13/37                                 & 35.1 & 0.97                    & --                  \\
SDSS J151020.06+554722.0 & 15 10 20.06   & +55 47 22.04  & 15/53                                 & 28.3 & 0.83                    & --                  \\
SDSS J152205.41+393441.3 & 15 22 05.41   & +39 34 40.71  & 4/61                                  & 6.6  & 1.43                    & 2.5                  \\
SDSS J164100.10+345452.7 & 16 41 00.11   & +34 54 52.68  & 2/71                                  & 2.8  & 0.46                    & 2.7                  \\ \hline

\end{tabular}
\tablefoot{
\tablefoottext{a}{This source has been detected also by LOFAR at a very low frequency of 150~MHz with a flux density of 3.6~mJy.}}
\label{tab:mh}
\end{table*}

\section{\emph{Fermi}--LAT data reduction}
\label{sec:fermidata}

Since the 37~GHz detections suggest the presence of jets, it is possible that these sources are also gamma-ray emitters. Therefore we obtained their \textit{Fermi} gamma-ray fluxes in the energy range 0.1--300~GeV by analysing the \textit{Fermi}--LAT data from 2008 August 04 to 2017 August 10 using the \textit{Fermi} Science Tools\footnote{http://fermi.gsfc.nasa.gov/ssc/data/analysis/documentation/Cicerone} version v10r0p5. We applied the data selection recommendation\footnote{https://fermi.gsfc.nasa.gov/ssc/data/analysis/documentation/Cicerone/Cicerone\_Data\_Exploration/Data\_preparation.html} for the Pass 8 data to select events in the source class, while excluding those with zenith angle $>90${\textdegree} to avoid contamination from photons coming from the Earth's limb. We used instrument response function P8R2\_SOURCE\_V6. The photons were extracted from a circular region centred on the source within a radius of 30{\textdegree}. We modelled all NLS1 galaxies using a PowerLaw2 spectral model\footnote{https://fermi.gsfc.nasa.gov/ssc/data/analysis/scitools/source\_models.html} by fixing their spectral index at -2.5 that was estimated from the average spectral indices of the detected NLS1s in the literature.

We used the binned likelihood method, and accounted for Galactic diffuse emission and isotropic background models using the templates 'gll\_iem\_v06.fits' and 'iso\_P8R2\_SOURCE\_V6\_v06.txt', provided in the Science Tools. Fluxes were obtained for the entire period assuming that for a detection TS \citep{1996mattox1} exceeds 25 ($\sim 5\sigma$). If TS  did not exceed this value, we computed upper limits using the profile likelihood method \citep{Rolke2005}. 

The position of the gamma-ray source was constrained using the \texttt{gtfindsrc} tool to coordinates RA 16h40m39.54s and Dec. +34d59m48.18s, separated by 0.10{\textdegree} from the radio counterpart. We also used an alternative method that localised the source by computing the likelihood TS with and without the source, which yielded a $\sim 5\sigma$ detection of the source at a separation of 0.12{\textdegree} from the radio counterpart.

\section{Black hole masses}
\label{sec:mbh}

For radio-silent sources, not included in \citet{2015jarvela1}, the black hole masses were estimated using the $FWHM$(H$\beta$) -- luminosity mass scaling relation \citep{2005greene1}

\begin{equation} M_{\text{BH}} = (4.4 \pm 0.2) \times 10^6 \bigg( \frac{\lambda L_{5100}}{10^{44} \text{ ergs s}^{-1}} \bigg)^{0.64 \pm 0.02} \bigg( \frac{\text{$FWHM$}(H\beta)}{10^3 \text{ km s}^{-1}} \bigg)^2 M_{\sun}
\label{eq:mbh} .\end{equation}

The $\lambda L_{\text{5100}}$ and $FWHM$(H$\beta$) values were taken from \citet{2006zhou1}.

We could not use methods based on $M_{\text{BH}}$ -- $\sigma_{\ast}$ (stellar velocity dispersion of the bulge) or $M_{\text{BH}}$ -- $M_{\text{bulge}}$ (the mass of the bulge) relation \citep{2009bentz1} because sufficient data are not available for most NLS1 galaxies. Our method does not take into account inclination effects caused by the geometry of the broad-line region and the viewing angle \citep{2011decarli2}, and might thus underestimate the black hole masses. On the other hand, because of the unknown amount of jet contamination to $\lambda L_{\text{5100}}$ this method overestimates the black hole masses of sources with relativistic jets \citep{2004wu1}.

\end{appendix}

\end{document}